# Non-thermal separation of electronic and structural orders in a persisting charge density wave

M. Porer[1,*], U. Leierseder[1], J.-M. Ménard[1], H. Dachraoui[2,†], L. Mouchliadis[3], I. E. Perakis[3], U. Heinzmann[2], J. Demsar[4], K. Rossnagel[5] & R. Huber[1,#]

[1]Department of Physics, University of Regensburg, 93040 Regensburg, Germany
[2]Molecular and Surface Physics, University of Bielefeld, 33615 Bielefeld, Germany
[3]Department of Physics, University of Crete & FORTH/IESL, Heraklion, Crete 71110, Greece
[4]Institute of Physics, Ilmenau University of Technology, 98684 Ilmenau, Germany
[5]Institute of Experimental and Applied Physics, University of Kiel, 24098 Kiel, Germany

[*]michael.porer@physik.uni-regensburg.de
[#]rupert.huber@physik.uni-regensburg.de

[†]Present address: Deutsches Elektronen-Synchrotron DESY, Notkestrasse 85, 22607 Hamburg, Germany

The simultaneous ordering of different degrees of freedom in complex materials undergoing spontaneous symmetry-breaking transitions often involves intricate couplings that have remained elusive in phenomena as wide ranging as stripe formation[1], unconventional superconductivity[1-7] or colossal magnetoresistance[1,8]. Ultrafast optical, x-ray and electron pulses can elucidate the microscopic interplay between these orders by probing the electronic and lattice dynamics separately[9-17], but a simultaneous direct observation of multiple orders on the femtosecond scale has been challenging. Here we show that ultrabroadband terahertz pulses can simultaneously trace the ultrafast evolution of coexisting lattice and electronic orders. For the example of a charge-density-wave (CDW) in 1$T$-TiSe$_2$, we demonstrate that two components of the CDW order parameter - excitonic correlations and a periodic lattice distortion (PLD) - respond very differently to 12-fs optical excitation. Even when the excitonic order of the CDW is quenched, the PLD can persist in a coherently excited state. This observation proves that excitonic correlations are not the sole driving force of the CDW transition in 1$T$-TiSe$_2$, and exemplifies the sort of profound insight that disentangling strongly coupled components of order parameters in the time domain may provide for the understanding of a broad class of phase transitions.

Spontaneous symmetry breaking gives rise to a new quantum ground state featuring characteristic low-energy elementary excitations[3,11,14,18-22]. Examples in complex solids include soft phonons[18], Cooper pairs[6-7,19], magnons[22], Josephson plasmons[14] and Higgs modes[19]. Ultrashort pulses in the terahertz (1 THz = $10^{12}$ Hz) range have been used to trace *electronic order* via direct coupling to such excitations[22,23]. We demonstrate that THz pulses may simultaneously also track the *crystalline order* during an ultrafast phase transition.

This idea is tested in a prominent reference system, $1T$-TiSe$_2$. Within the family of layered transition-metal dichalcogenides, this material has attracted special attention: Upon cooling below $T_c \approx 200$ K, it undergoes a transition into a commensurate CDW accompanied by the formation of a structural (2×2×2) superlattice[21] (Fig. 1a). In its high-temperature phase, TiSe$_2$ is a semimetal[20] with electron and hole pockets at the L and Γ points of the Brillouin zone, respectively[15,24] (Fig. 1b). The spatial reconstruction due to the CDW maps these two points on top of each other and leads to the partial opening of an electronic energy gap as well as a dramatic reduction of the density of free charge carriers[20] (Fig. 1b). Superconductivity emerges when the CDW is suppressed, e.g. by Cu intercalation[7] or pressure[25]. This discovery as well as novel chiral properties[26] have intensified the interest in the nature of the CDW in $1T$-TiSe$_2$. Yet, the microscopic mechanisms remain elusive[24,27-29]. A first hypothesis assumes electron-phonon coupling based on a Jahn-Teller effect as the driving force[27]. A competing model suggests that the transition is purely electronically driven[24,28]. Coulomb attraction may render the system unstable against the formation of excitons between the electron- and hole-like Fermi pockets, leading to lattice deformation with the corresponding wave vector. Combinations of the two scenarios have also been proposed[29]. Time-resolved x-ray diffraction[16] and photoemission[10,15] experiments have separately tracked the dynamics of either structural or electronic orders.

Evidence for both excitonic and phononic contributions was obtained in this way, leaving a controversial picture.

Here we disentangle the two coupled components of the CDW order parameter by simultaneously tracing the ultrafast THz response of PLD-related phonons and electronic conductivity while a femtosecond pulse selectively melts the excitonic order. Our data reveal a transient phase in which the PLD persists in the absence of excitonic correlations. A quantum-mechanical theory[29] corroborates our conclusions.

In TiSe$_2$, the transition to the CDW ordered phase modifies the low-frequency optical response in three distinct ways: (i) The CDW-induced energy gap introduces a broad single-particle Se4$p$-Ti3$d$ interband resonance in the optical conductivity centred at 0.4 eV with its low-energy tail extending down to 0.1 eV[20] (Figs. S3 and S9). (ii) Electronic order manifests itself most accessibly in the energy loss function -Im(1/$\varepsilon$)[20]. At room temperature, collective plasma oscillation of unbound electrons and holes causes a broad maximum at a central energy of 145 meV, with a linewidth set by the scattering time $\tau_N$ = 40 fs. Below the CDW phase transition, $T < T_c$, electronic correlations reduce the density of free carriers $n$ by one order of magnitude, leaving behind a dilute electron gas with $n_{CDW}$ = 9 × 10$^{19}$ cm$^{-3}$. Thus the screened plasmon pole shifts to $\hbar\omega_{p,CDW}$ = 45 meV (Fig. 1c). The reduced phase space for scattering[20] in the ordered state narrows the linewidth (FWHM) to 7 meV ($\tau_{CDW}$ = 0.9 ps). (iii) The PLD also changes the THz response in a characteristic way (Fig. 1d). Above $T_c$, we observe a single TO phonon resonance at 17 meV. Below $T_c$, back-folding of the uppermost acoustic branch from the L to the Γ point[21] yields an additional IR-active in-plane mode at 19 meV. The weaker peak at 22 meV likely originates from a folded optical branch at the M point[21].

We exploit these spectral fingerprints to trace the ultrafast dynamics of the electronic order and the PLD. Therefore, we perturb an 80 nm thin film of stoichiometric 1$T$-TiSe$_2$ by 12-fs near-infrared pulses (see Methods). Co-propagating multi-THz transients probe the low-energy response, after a variable delay time $t_D$. Electro-optic sampling of the transmitted THz waveforms allows us to obtain the instantaneous complex dielectric response $\varepsilon(\omega,t_D)$ (Ref. 30 and Supplementary Information). Figure 2a shows the measured spectra of $-\text{Im}(1/\varepsilon)$ as a function of $t_D$, for an absorbed pump fluence $\Phi = 20$ µJ/cm$^2$ at $T = 10$ K. At negative delays, the system is in thermal equilibrium characterised by the sharp plasmon pole at $\hbar\omega_{p,\text{CDW}} = 45$ meV. Upon photoexcitation, this peak rapidly blue-shifts and then recovers within 1.5 ps. The resonance remains narrow during the entire dynamics, indicating that the charge order stays largely intact while the pump injects electron-hole pairs. In addition, we observe a transient increase of spectral weight on the low-energy side of the single particle resonance associated to the CDW energy gap (Fig. S4). An elevated fluence of $\Phi = 330$ µJ/cm$^2$ triggers qualitatively different dynamics (Fig. 2b). The plasma resonance shifts above our measurement window and stays strongly broadened during the subsequent relaxation. As shown below, these signatures reflect the complete destruction of the excitonic order. After a delay of $t_D \approx 2.5$ ps, a sharp plasmon pole re-emerges at its equilibrium position. Notably, the spectral signature of the CDW energy gap persists during this process (Fig. S9).

For a quantitative analysis of the plasma response, we utilise both real and imaginary parts of $1/\varepsilon(\omega,t_D)$ to extract the instantaneous Drude parameters $\tau$ and $\omega_p$ (Fig. S2) as well as the corresponding transient free carrier density $n$. The photoinduced density change $\Delta n$ (Fig. 2c) rises within 100 fs (Fig. S6) followed by an exponential decay with a time constant of 0.5 ps. This holds for all fluences tested. In contrast, the temporal profile of $\tau$ depends critically on $\Phi$ (Fig. 2d). Starting at $\tau_{CDW} = 0.9$ ps, $\tau$ drops rapidly after photoexcitation and recovers on a fluence-dependent ps timescale (Fig. S8). For $\Phi \leq 20$ μJ/cm$^2$, $\tau$ remains distinctly above $\tau_N$, indicating that the electronic order is not completely destroyed. In the case of $\Phi \geq 60$ μJ/cm$^2$, $\tau$ abruptly decreases by a factor of ~20. The system reaches normal state conductivity with $\tau_N = 40$ fs and remains there for up to 2.5 ps (depending on $\Phi$) before it starts to recover.

To assess the role of the excitation density more quantitatively, we evaluate $n$ and $\tau$ as a function of $\Phi$, at fixed $t_D = 0.2$ ps (Figs. 2e and 2f). When exciting the CDW phase ($T = 10$ K), $n$ grows *superlinearly* with $\Phi$, changes its curvature at a threshold fluence $\Phi_{th} \approx 40$ μJ/cm$^2$, and approaches a linear dependence for $\Phi > 140$ μJ/cm$^2$ (Fig. 2e). Simultaneously, $\tau$ decreases with $\Phi$, reaches the normal phase limit, and then levels off for the highest pump fluences (Fig. 2f). The fluence required to switch to normal state conductivity coincides with the inflection point $\Phi_{th}$ of the function $n(\Phi)$ (Fig. 2e). Excitation below $\Phi_{th}$ leaves the charge order partly intact, since $\tau$ remains above its normal state value[20]. In this regime the electronic system cannot be described by a thermalised state since the combination of large $n$ with $\tau \gg 40$ fs is not observed at any equilibrium temperature (Fig. S10). In an analogous study starting from the normal state, the superlinear increase of $n$ is absent (Fig. S7).

The observed superlinear response is consistent with the picture of charge-transfer excitonic correlations[29]. The optical pump generates primary electron-hole pairs with densities scaling linearly with $\Phi$. Their excess energy is relaxed via cascaded electron-electron scattering, which multiplies the quasi-particle density. For low excitation density ($\Phi < \Phi_{th}$), the photogenerated carriers screen Coulomb interactions and thus reduce the excitonic binding potential. As a consequence, part of the bound pairs breaks into free carriers that further enhance screening. This scenario is supported by a quantum-mechanical model calculation (Fig. S13a) and explains the superlinear increase of $n$. As soon as the carrier density exceeds a critical value $n_c$, excitonic correlations can no longer exist, τ assumes its normal state value (Fig. 2f) and the superlinear scaling of $n$ stalls (Fig. 2e). Upon full suppression of Coulomb correlations, only the linear dependence of $n$ on Φ remains (dashed curve in Fig. 2e). From the inflection point of $n(\Phi)$ we find $n_c \approx 4\times10^{20}$ cm$^{-3}$ at $T = 10$ K. This situation indeed corresponds to a Thomas-Fermi screening length comparable to the CDW wavelength (Supplementary Information). The excitonic order recovers faster from a perturbed state than from complete suppression (Fig. S8) and only sets in after $n$ has decreased below $n_c$.

While the plasmon pole witnesses the electronic order, the polarizability of back-folded phonon branches allows us to simultaneously follow the PLD[21]. Figure 3a summarises the phonon response of the photoexcited sample for $T = 10$ K and $\Phi = 330$ μJ/cm$^2$, i.e. after complete quenching of excitonic correlations. The back-folded mode at ℏω = 19 meV (see also Fig. 1) remains virtually unaffected at all times $t_D$. This unexpected observation demonstrates that the PLD remains stable, even though Φ is sufficiently high to keep the excitonic order molten for several ps (Fig. 2). If electronic interactions were the main driving force of the PLD, this would be enough time for the lattice to relax into its undistorted state. Yet this is not observed. In order to achieve a melt-down of the PLD with the maximal excitation density available the sample has to be heated to $T = 150$ K (Fig. 3b). The fluence required to optically melt the PLD roughly corresponds to half of the energy required to heat the excited volume to $T = T_c$ (Supplementary Information). On the other hand, the electronic order is quenched with fluences lower by one order of magnitude.

Both the remaining PLD and the CDW gap demonstrate that the CDW phase can exist without excitonic order. Nevertheless the lattice and electronic orders do couple: Photoexcitation induces coherent modulations of the CDW gap superimposed to an exponential recovery dynamics (Figs. S4 and S5). The oscillating component of the photoinduced THz transmission (Fig. 4a) exhibits a frequency of 3.4 THz, for $\Phi = 7$ µJ/cm$^2$ (Fig. 4b), which is characteristic of the well-known A$_{1g}$ CDW amplitude mode[16]. While the oscillation is still observed for excitation densities above $\Phi_{th}$, its amplitude increases with $\Phi$ only up to $\Phi_{th}$ (Fig. 4c). This behaviour is indeed expected when the CDW gap is jointly formed by excitonic correlations and Jahn-Teller effects. Following impulsive weakening of the excitonic order, the remaining PLD oscillates around a slightly relaxed potential energy minimum. This coherent dynamics modulates the Jahn-Teller component of the CDW gap. Following a complete quench of excitonic correlations, the new equilibrium position is stabilised by Jahn-Teller effects and the oscillation amplitude, given by the pump-induced shift of the equilibrium position, remains at its maximum level. This interpretation is supported by our theory (Supplementary Information) predicting a reduced, but finite PLD when the excitonic order is quenched. The precise traces of the coherent oscillations are recorded at different excitation densities (Fig. 4a). The initial phase of the coherent oscillation is constant for all fluences. Yet, its frequency slightly decreases when excited with $\Phi = 67$ µJ/cm$^2$ as compared to $\Phi = 3$ µJ/cm$^2$. The change in frequency eventually leads to a phase retardation of ~$\pi/2$ after three oscillation periods at $t_D = 1$ ps (black arrow in Fig. 4a). Since $\Phi = 67$ µJ/cm$^2$ quenches excitonic correlations within the initial ~1 ps time window (Fig. 2d), the restoring force in this case can only be given by a remaining Jahn-Teller mechanism. Indeed, the pump-induced softening of the A$_{1g}$ amplitude mode is reproduced by our theoretical model and suggests a cooperative coupling between excitonic correlations and Jahn-Teller effects (Supplementary Information).

For the case of 1$T$-TiSe$_2$, our results confirm a scenario[29] where the CDW order is composed of Jahn-Teller effects and excitonic ordering. By rapidly quenching the excitonic component, we show that the remaining Jahn-Teller-like CDW can exist in a metastable non-thermal phase without excitonic correlations. Our idea to disentangle the ultrafast dynamics of structural and electronic components of the order parameter via probing back-folded phonon branches together with a purely electronic response should be universally applicable to a broad variety of materials. Similarly exciting insights are anticipated for materials as diverse as VO$_2$ [17], iron pnictides[11], cuprates[2,4,5,14] and other unconventional superconductors[1,6], which have all proved hard to understand due to the strong coupling between different orders.

**Methods**

The system under study is a bulk single-crystal of stoichiometric 1$T$-TiSe$_2$ grown by iodine vapour transport. We contact the sample featuring a lateral size of (0.2×0.2) mm$^2$ by van der Waals bonding with a CVD-grown diamond window. Repeated exfoliation allows us to reduce the thickness of the TiSe$_2$ film to a value of $L$ = 80 nm, as confirmed by atomic force microscopy and optical transmission. For our experiments, intense 12-fs light pulses centred at a photon energy of 1.55 eV are derived from a high-repetition rate Ti:sapphire amplifier system. A first portion of the laser output photoexcites the sample. Optical rectification of another part in a 15 μm thick GaSe crystal generates single-cycle THz probe transients covering more than two optical octaves with photon energies ranging between 35 and 130 meV. Alternatively, optical rectification in a GaP emitter provides field transients covering the energy window between 3 and 30 meV.

**Acknowledgements**

We thank A. Pashkin and R. Bratschitsch for helpful discussions as well as M. Furthmeier, C. Gradl, K. Groh, T. Riedel and C. Sohrt for technical assistance. Support by the European Research Council via ERC grant 305003 (QUANTUMsubCYCLE) is acknowledged. I.E.P. and L.M. were supported by the European Union's Seventh Framework Programme (FP7-REGPOT-2012-2013-1) under grant agreement N$^o$ 316165.

**Author contributions**

M.P., H.D., U.H. and R.H. planned the project; M.P., U.L. and J.M.M. performed terahertz measurements; M.P., J.D., U.H., J.M.M., K.R. and R.H. analysed data; K.R. and U.H. provided bulk samples; M.P. prepared thin film samples; M.P., L.M. and I.E.P. elaborated the theoretical model; M.P., J.D., K.R., I.E.P. and R.H. wrote the paper. All authors contributed to discussions and gave comments on the manuscript.

**The authors declare no competing financial interests.**

**Figure 1 | CDW phase transition in 1$T$-TiSe$_2$ and its low-energy spectral fingerprint. a**, The periodic lattice distortion associated with the CDW (space group: $P\bar{3}c1$) causes a doubling of the unit cell of the undistorted high-temperature structure (pale spheres in the background, space group: $P\bar{3}m1$), in all three spatial dimensions. The structure is viewed along the c-axis. The dashed line frames the unit cell in the CDW phase. **b**, Schematic electronic band structure above (upper panel) and below (lower panel) $T_c$. Back-folding of the 3$d$-like conduction bands at the L point onto the zone centred 4$p$-like valence bands by $q_{CDW}$ yields a gapped structure[24] (blue). **c**, Mid-infrared energy loss function $-\text{Im}(1/\varepsilon)$ in the normal ($T$ = 300 K, red curve) and the CDW phase ($T$ = 10 K, blue curve) according to the model function of Ref. 20. The spectrum at $T$ = 10 K results from a numerical adaption to the infrared reflectance of our sample (see Fig. S1). **d**, Imaginary part of the dielectric function measured by THz time domain spectroscopy, revealing characteristic in-plane polarised TO-phonons in the normal (red curves) and the CDW phase (blue curves). Arrows indicate phonon resonances that are only observable in the presence of the PLD.

**Figure 2 | Ultrafast photoinduced dynamics of the mid-infrared electronic response. a,b**, Two-dimensional colour maps of the transient energy loss function $-\text{Im}(1/\varepsilon)$ versus photon energy and delay time at $T$ = 10 K, following near-infrared excitation with $\Phi$ = 20 µJ/cm$^2$ (**a**) and $\Phi$ = 330 µJ/cm$^2$ (**b**). Dashed lines are guides to the eye indicating the centre frequency of the plasmon pole. **c**, Transient change in the free carrier density $n$ and **d**, the carrier scattering time $\tau$ extracted from two-dimensional THz time domain spectroscopy at various pump fluences. The change in $n$ measured for $\Phi$ = 7 µJ/cm$^2$ is upscaled by a factor of 120 for quantitative comparison with the situation at $\Phi$ = 170 µJ/cm$^2$ (**c**). Horizontal dotted line in (**d**): $\tau = \tau_N$. **e**, Number of free carriers $n$ and **f**, carrier scattering time $\tau$ as a function of $\Phi$ at $t_D$ = 0.2 ps

measured at $T = 10$ K. Vertical shaded lines indicate the threshold fluence $\Phi_{th} \approx 40$ µJ/cm$^2$ required for transient suppression of the electronic correlations. Horizontal dotted line in (**e**): Critical density $n_c = 4 \times 10^{20}$ cm$^{-3}$. Grey dashed line: Linear regression of $n(\Phi)$ in the high fluence regime ($\Phi > 0.14$ mJ/cm$^2$). Horizontal dotted line in (**f**): $\tau = \tau_N$.

**Figure 3 | THz phonon spectrum during ultrafast melting of the electronic order. a**, Imaginary part of the THz dielectric function at select delay times $t_D$ after optical excitation with $\Phi = 0.33$ mJ/cm$^2$ and $T = 10$ K. The phonon resonance originating from the back-folded acoustic branch remains visible (black arrows) at all times, proving the persistence of the PLD. The elevated conductivity of photoexcited charge carriers found at a delay time of 1.5 ps (compare Fig. 2b) accounts for the increased background compared to the other spectra. **b**, Same measurement for $T = 150$ K. The effective suppression of the PLD-related phonon (red crosses) for $t_D = 0$ and 3 ps attests to a complete melting of the PLD.

**Figure 4 | CDW amplitude oscillations following a perturbation of the electronic order. a,** Oscillatory component $\Delta A_{osc}(t_D)$ of the pump-induced THz transmission change $\Delta A(t_D)$ recorded at $T = 10$ K and various $\Phi$. $\Delta A$ is spectrally integrated over an energy window from 40 meV to 140 meV. Vertical grey lines mark the maxima of the waveform recorded with $\Phi = 3$ µJ/cm$^2$. The black vertical line and the black arrow indicate the time difference required for three full oscillation cycles during persistent ($\Phi = 3$ µJ/cm$^2$, blue line) and suppressed ($\Phi = 67$ µJ/cm$^2$, purple line) electronic order. **b,** A representatively selected frequency spectrum of the oscillations observed with $\Phi = 7$ µJ/cm$^2$ identifies the well-known $A_{1g}$ CDW amplitude mode[16]. **c,** Maximum amplitude of $\Delta A_{osc}$ as a function of $\Phi$. The slope flattens at a fluence coinciding with $\Phi_{th}$ (shaded vertical line).

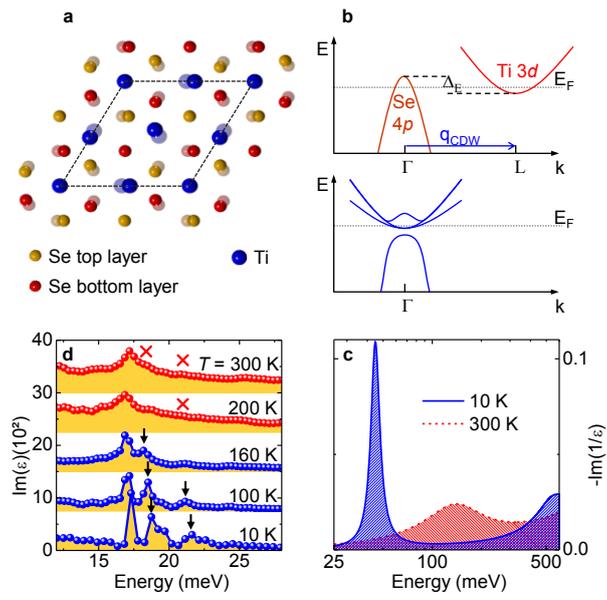

Porer et al., Figure 1

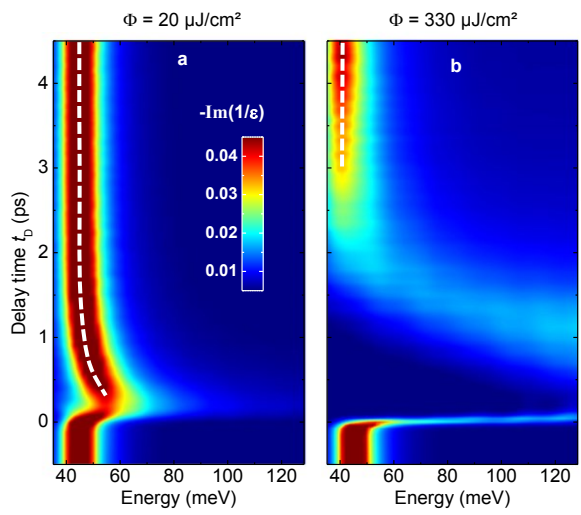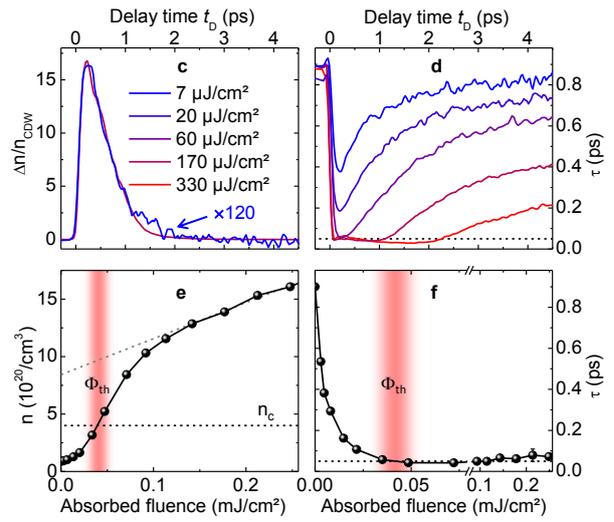

Porer et al., Figure 2

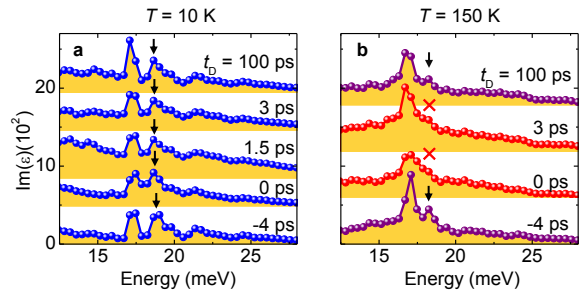

Porer et al., Figure 3

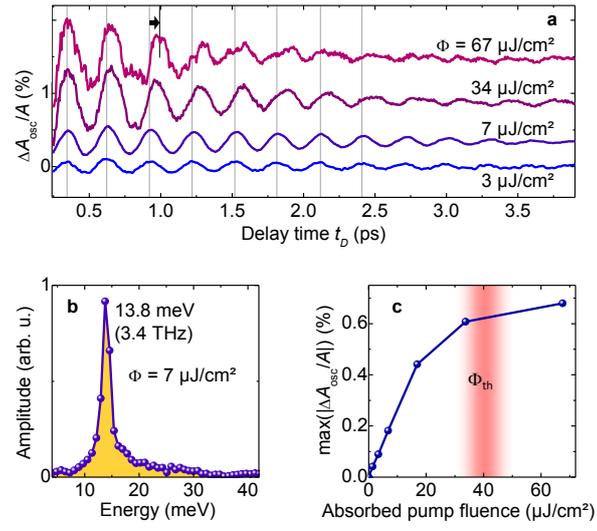

Porer et al., Figure 4